\documentclass[aps,onecolumn,twoside]{revtex4}
\usepackage{graphicx}  % needed for figures

\newcommand{\Bbar}{\,\overline{\!B}}

\newcommand{\bsb}{\ensuremath{\Bbar{}_s}}

\newcommand{\dm}{\ensuremath{\Delta M}}
\newcommand{\dg}{\ensuremath{\Delta \Gamma}}

\newcommand{\ov}[1]{\overline{#1}}

\newcommand{\eq}[1]{(\ref{#1})}

\newcommand{\etal}{\emph{et al}}

\newcommand{\fb}  {\mbox{fb}^{-1}}

\newcommand{\order}[1]{\mathcal{O}( #1 )}
\newcommand{\asl} {A_{SL}}

\newcommand{\adcpv}{A_{CP}(B^+ \to J/\psi K^+)}

\newcommand{\jp}{J/\psi}
\newcommand{\jpsi}{J/\psi}

\newcommand{\fig}[1]{Fig.~\ref{#1}}

\begin{document}

\title{CP asymmetries at D0}
\author{K Holubyev}
\affiliation{Lancaster University, UK}
\email{holubyev@fnal.gov}

\begin{abstract}
Using two independent measurements of the semileptonic CP asymmetry
in the $B_s$ system, we constrain the CP-violating phase
of the $B_s$ system to be  $\phi_s = -0.70^{+0.47}_{-0.39}$.
The data sample corresponds to an integrated luminosity of 1.1 $\fb$
accumulated with D0 detector at the Fermilab Tevatron Collider.
We also measure the direct CP violating asymmetry
in the decay $B^+ \to \jp K^+$ to be 
$\adcpv = +0.0067 \pm 0.0074$(stat)$\pm0.0026$(syst).
The data corresponds to an integrated luminosity of 1.6 $\fb$.
\end{abstract}

\maketitle

\section{Semileptonic CP asymmetry in the $B_s$ system}

% Introduction in Bs and phi_s

In the Standard Model (SM), the light (L) and heavy (H) 
mass eigenstates of the mixed $B_s^0$ 
system are expected to have sizable mass and decay width differences:
$\dm_s \equiv M_H - M_L$ and $\dg_s \equiv \Gamma_L - \Gamma_H$. The two
mass eigenstates are expected to be alsmost pure CP eigenstates. 
The CP-violating mixing phase is predicted \cite{lenz} to be 
$\phi_s = (4.2 \pm 1.4) \times 10^{-3}$. New phenomena may alter $\phi_s$
leading to a reduction of the observed $\dg_s$ compared to the SM prediction 
$\dg_s^{SM}$: $\dg_s = \dg_s^{SM} \times |\cos \phi_s|$. While 
$B_s^0 - \bsb^0$ oscillations have been detected \cite{comb5} and the mass
difference has recently been measured \cite{cdf}, the CP-violating phase
remains unknown. 

% base - Bs to jp phi data

Both CP-violating phase $\phi_s$ and decay width difference $\dg_s$ 
of the $B_s$ system were
for the first time directly constrained at D0
from the fit to the time-dependent angular distribution
of the decay products in the decay sequence $B_s^0 \to
\jp \phi$, $\jp \to \mu \mu$, $\phi \to K^+ K^-$ \cite{jpphi}.
The result remained 4-fold ambiguous due to undefined
CP-conserving strong phases.
The semileptonic asymmetry in the $B_s$-system, 
which is in general defined as
\begin{eqnarray}
\asl^s = \frac{N(\bsb^0 \to l^+ X) - N(B_s^0 \to l^- X)}{N(\bsb^0 \to l^+ X) + N(B_s^0 \to l^- X)},
\label{asl-def}
\end{eqnarray}
is related to both $\phi_s$ and $\dg_s$ via 
\begin{eqnarray}
\asl^s = \frac{\dg_s}{\dm_s}\tan \phi_s.
\label{asl-phi}
\end{eqnarray}
Its measurement gives independent access to $\phi_s$, 
both resolving the mentioned ambiguity and 
adding statistics to the measurement.

\begin{figure}
\begin{center}
\includegraphics[width=0.7\textwidth]{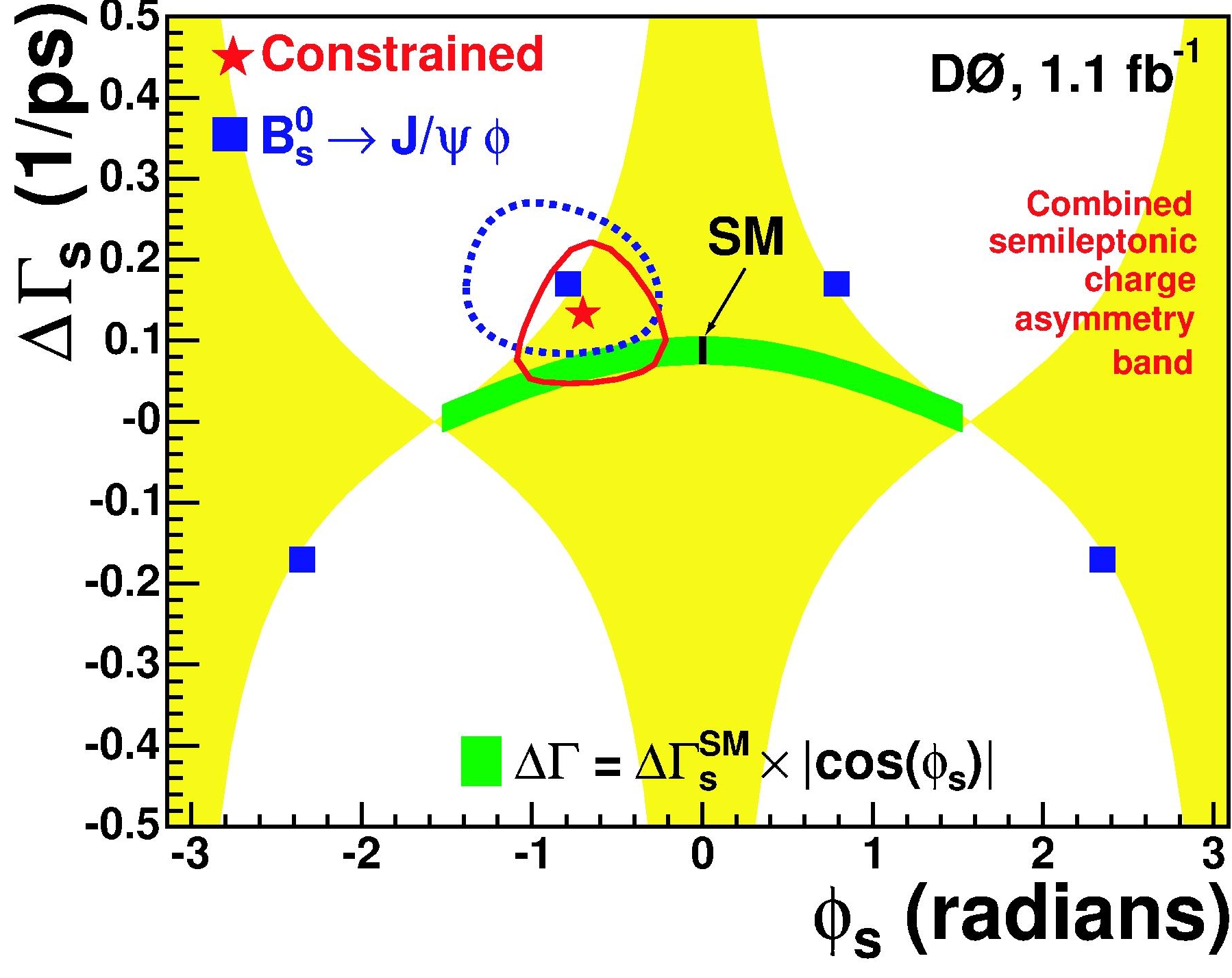}
\caption{The error ellipse ($\Delta \ln (\mathcal{L})=0.5$) in the plane
($\dg_s$, $\phi_s$) for the fit to the $B_s^0 \to \jp \phi$ data
(dashed line) and for the fit with the constraint from the two
D0 measurements of the charge asymmetry $\asl^s$ in semileptonic $B_s^0$ decays
(solid line). The central values of four solutions of the unconstrained
fit are indicated by squares. Also shown is the band representing 
the relation $\dg_s = \dg_s^{SM} \times |\cos \phi_s|$
with $\dg_s^{SM} = 0.088 \pm 0.017$ ps$^{-1}$ \cite{lenz} (dark shade)
and the area corresponding to 
$\dg_s \cdot \tan \phi_s = 0.02 \pm 0.16$ ps$^{-1}$ \cite{cdf} (light shade).}
\label{fig-comb}
\end{center}
\end{figure}

Recently we at D0 accessed the semileptonic asymmetry $\asl^s$
indirectly, by measuring the dimuon asymmetry
in the inclusive dimuon sample \cite{bruce},
and directly, by measuring the untagged asymmetry in the exclusive
sample of events consistent with the decay $B_s^0 \to \mu \nu_{\mu} D_s$, 
$D_s \to \phi \pi$ \cite{asl}.
The combination of the two results gives the best estimate
of the charge asymmetry in semileptonic $B_s^0$ decays:
$\asl^s = 0.0001 \pm 0.0090$ \cite{comb}. Using \eq{asl-phi}
and the result $\dm_s = 17.8 \pm 0.1$ ps$^{-1}$ from CDF experiment
\cite{cdf} we obtained $\dg_s \cdot \tan \phi_s = \asl^s \cdot \dm_s=
0.02 \pm 0.16$ ps$^{-1}$. Using this constraint we repeated the fit 
to the $B_s^0 \to \jp \phi$ data. 
In \fig{fig-comb} we show the likelihood
contours in $\dg_s$ vs $\phi_s$ plane without (dashed line)
and with (solid line) the constraint from
the measurements of the semileptonic asymmetry $\asl^s$ in the $B_s^0$ decays. 
The contours indicate error
ellipses, $\Delta \ln(\mathcal{L}) = 0.5$, corresponding
to the confidence level of 39\%.

Finally, from the fit likelihood profile we found for $\phi_s < 0$
the decay width difference and the CP-violating phase in the $B_s$-system
to be $\dg_s = 0.13 \pm 0.09$ ps$^{-1}$, 
$\phi_s = -0.70^{+0.47}_{-0.39}$.
The measurement uncertainty is dominated  by limited statistics.
The systematic uncertainties include a variation of the background model
in the analysis of the decay $B_s^0 \to \jp \phi$, detector
acceptance, and sensitivity to the details of track and vertex
reconstruction. The results are consistent with the SM predictions \cite{lenz}.

% Results of comb

\section{Direct CP violation in the decay $B^+ \to \jp K^+$}

A direct CP asymmetry in the decay $B^+ \to \jp K^+$, $\adcpv$,
has recently been measured at D0:
\begin{eqnarray}
\adcpv = \frac{N(B^- \to \jp K^-) - N(B^+ \to \jp K^+)}
              {N(B^- \to \jp K^-) + N(B^+ \to \jp K^+)}
\label{a-dcpv}
\end{eqnarray} 
This decay proceeds via $b \to c \ov{c} s$ transition which is 
predominantly tree level. The SM gives the order
of magnitude estimate $\adcpv = \order{0.003}$ \cite{hou}, which in the realistic
New Physics (NP) models can be enhanced to 0.01 or higher \cite{hou}.

% Results for DCPV

\begin{figure}
\begin{center}
\includegraphics[width=0.7\textwidth]{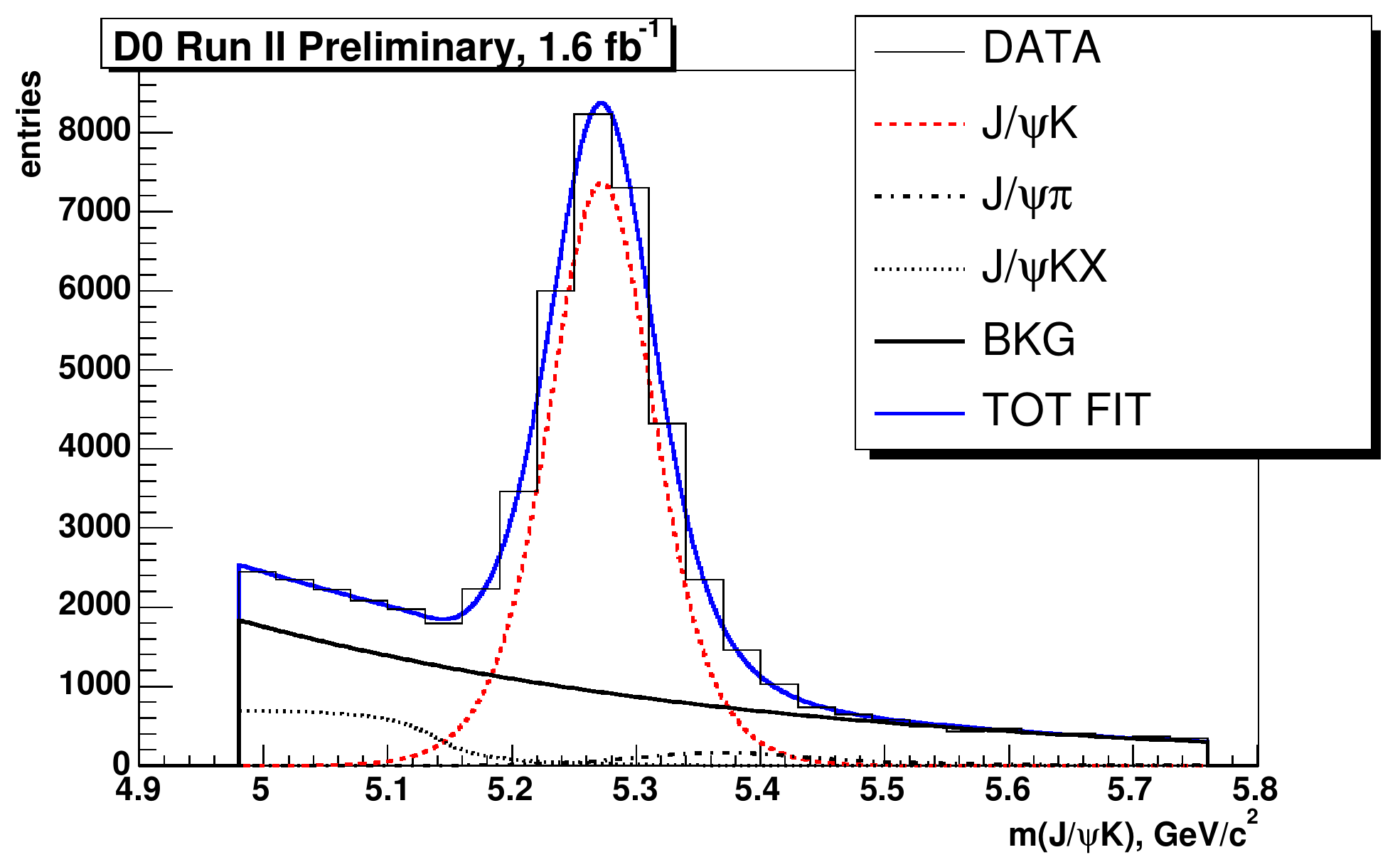}
\caption{Result from the unbinned fit of the invariant mass distribution
of the $\jp K$ system in the $B^+ \to \jp K^+$ decay  and its charge conjugate.}
\label{mjpk-fit}
\end{center}
\end{figure}

The events
consistent with the decay chain $B^+ \to \jp K^+$,
$\jp \to \mu^+ \mu^-$ and its charge conjugate were selected.
The $\jp K$ mass peak was modeled using
unbinned likelihood fit to the sum
of contributions from $B \to \jpsi K$, $B \to \jpsi \pi$, and
$B \to \jpsi K^{*}$ decays, as well as combinatorial background, 
see \fig{mjpk-fit}.

The systematic shift from the detector-induced asymmertries was
accounted for in the detector model (for the first time applied in \cite{bruce}), 
which expresses the number of
signal events with the kaon charge $q$, 
the sign of the kaon pseudorapidity $\gamma$,
and the solenoid polarity at which the event
was recorded $\beta$ in terms of the kaon charge asymmetry $A$ and various
detector asymmetries $A_i$ (see section IV of \cite{bruce}
for the explanation of $N$, $\epsilon^{\beta}$, and different $A_i$):
\begin{eqnarray}
n_{q}^{\beta \gamma} & = & \frac{1}{4} N \epsilon^{\beta}(1+qA)(1+q \gamma A_{fb})
(1+\gamma A_{det}) (1+q \beta \gamma A_{q \beta \gamma}) 
(1+q \beta A_{q \beta}) (1+\beta \gamma A_{\beta\gamma}).
\label{detpar}
\end{eqnarray}
The initial data sample of \fig{mjpk-fit}
was divided into subsamples corresponding to eight possible
combinations of $\beta$, $\gamma$, and $q$, in each subsample 
the unbinned fit was performed
to find the number of events
in the $\jp K$ peak, $n_{q}^{\beta \gamma}$, and the system \eq{detpar}
was solved for all asymmetries.

A systematic shift from charge
asymmetric kaon interactions with detector material
was estimated from data
by comparing the extlusive decay
$c \to D^{*+} \to D^0 \pi^+$, $D^0 \to \mu^+ \nu_{\mu} K^-$
and its charge conjugate.
To account for the momentum dependence
of the kaon cross-section \cite{pdg},
the kaon asymmetry in the $D^*$ sample was measured
in kaon momentum bins to convolve it
with the PDF of the kaon momentum in the $\jp K$ sample.

Finally, after subtracting kaon asymmetry, we obtained 
$\adcpv = +0.0067 \pm 0.0074$(stat)$\pm0.0026$(syst),
which is consistent with the PDG-2007 world average, 
$A_{CP}(B^+ \to \jp K^+)=+0.015 \pm 0.017$ \cite{pdg},
but has a factor of two better precision, thus providing
the most stringent bounds for new models predicting
large values of $\adcpv$.
The measurement uncertainty is mainly due to limited statistics.
Systematic uncertainty is largely dominated by the variation
of the $\jp K$ mass peak model.

%\section*{References}

\end{document}